\documentclass[prb,superscriptaddress,showpacs,twocolumn]{revtex4}%
\usepackage[english]{babel}

\usepackage{amsmath}
\usepackage{amsfonts}
\usepackage{amssymb}

\usepackage{graphicx}
\graphicspath{{images/}}

\usepackage{braket}
\usepackage{epstopdf}
\usepackage{color}
\usepackage{bm}%
\usepackage{hyperref}
\hypersetup{
	colorlinks   = true, 
	urlcolor     = blue, 
	linkcolor    = blue, 
	citecolor   = blue 
}

\newcommand{\be}{\begin{equation}}
\newcommand{\ee}{\end{equation}}
\newcommand{\bea}{\begin{eqnarray}}
\newcommand{\eea}{\end{eqnarray}}
\newcommand\ba{\begin{align}}
\newcommand\ea{\end{align}}
\def\nn{\nonumber}

\newcommand{\appropto}{\mathrel{\vcenter{
			\offinterlineskip\halign{\hfil$##$\cr
				\propto\cr\noalign{\kern2pt}\sim\cr\noalign{\kern-2pt}}}}}

\renewcommand{\vec}[1]{{\bf #1}}
\renewcommand{\hat}[1]{{\widehat #1}}

\renewcommand{\Re}{{\rm \, Re\,}}
\renewcommand{\Im}{{\rm \, Im\,}}

\newcommand{\Tr}{{\rm Tr\,}}

\newcommand{\p}{\partial}

\newcommand{\D}{{\cal D}}

\newcommand{\lb}{\left[}
\newcommand{\rb}{\right]}
\newcommand{\lp}{\left(}
\newcommand{\rp}{\right)}

\begin{document}
	
\title{ Ultrasonic attenuation via energy diffusion channel in  disordered conductors}
\author{Alexander Shtyk}
\affiliation{ L. D. Landau Institute for Theoretical Physics, Chernogolovka, Moscow
	region, Russia}
\affiliation{ Moscow Institute of Physics and Technology, Dolgoprudny, Moscow region,
	Russia}
\affiliation{Department of Physics, Harvard University, Cambridge, Massachusetts 02138, USA}
\author{Mikhail Feigel'man}
\affiliation{ L. D. Landau Institute for Theoretical Physics, Chernogolovka, Moscow
	region, Russia}
\affiliation{ Moscow Institute of Physics and Technology, Dolgoprudny, Moscow region,
	Russia}
\date{\today}

\begin{abstract}
We predict an existence of an new dissipation channel leading to attenuation of ultrasound
in disordered conductors and superconductors with perfect electroneutrality. 
It is due to slow diffusion of thermal energy.
We show that in doped silicon ultrasound attenuation may be enhanced by a factor about 100.
Similar effect is also studied for s-wave and d-wave superconductors. 
The latter  case is applied to  BSCCO family  where strong enhancement of ultrasound attenuation is predicted.
For usual s-wave superconductors new dissipation channel  might be important for very low-electron-density 
materials near the BCS-BEC crossover.

\end{abstract}

\maketitle

\section{Introduction and model.}

Ultrasonic attenuation in metals have already been studied for a long time\cite{pippard, akhiezer, blount, tsuneto, schmid} and may seem to be fully understood. The ratio of ultrasound attenuation rate $\alpha(\omega)$ to the sound frequency $\omega$  in clean metal is small due to adiabatic parameter $m k_F^3/\rho_m \ll 1$\, where $m$ is the electron mass, $k_F$ is the Fermi wave vector and $\rho_m$ is the density of the material. The simplest model for electron-phonon interaction is the scalar vertex Frohlich model with an extension due to  Migdal \cite{frohlich,migdal}. This model is applicable for clean metals. However, it is well known that the Frohlich model is not adequate when phonon wavelength $2\pi/q$ exceeds elastic electron mean free path $l$. In this \emph{dirty} limit, when $ql\ll1$, the conventional theory of electron-phonon interaction in a disordered conductors  leads to  the Pippard ineffectiveness condition(PIC) that tells that the ultrasonic attenuation at small wave vectors is suppressed by a factor $ql\ll1$ \cite{schmid}.

The  PIC phenomenon results from  strong Coulomb interaction that prohibits any charge disbalance in the system (perfect screening condition). Still some mechanisms were found recently which increase inelastic electron-phonon rate even in presence of perfect screening. The first of the them is present in  a multiband electronic spectrum \cite{multiband,us}, the second one is realized when impurities do not quite follow a motion of a lattice~\cite{sergeev_mitin}. Still another possibility is related with a deviations from the perfect screening condition that become prominent in a strongly disordered conductors, where $k_Fl$ is not very large~\cite{us,sergeev_2}.

In the present  paper we show that even in the simplest case of one-band electron spectrum and a perfect screening, the conventional PIC theory might still be unsufficient. The reason  is that electron liquid possesses an intrinsic  diffusion mode  that is present even under the condition of strict electroneutrality. It is an \emph{energy diffusion mode}  that  plays a role  similar to the spin polarization mode studied in Ref.~\cite{us}. Coupling of phonons to this diffusion mode results in an additional attenuation of longitudinal sound waves.

\begin{figure}[t]
	\center{\includegraphics[width=1\linewidth]{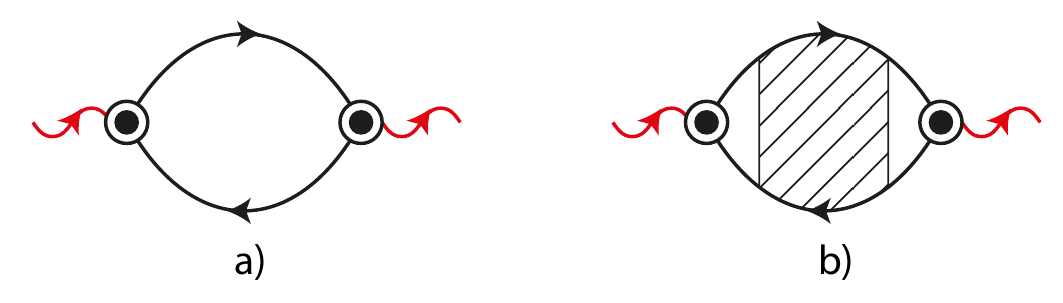}} 
	\caption{(Color online) Phonon self energy for (a) local processes and (b) diffusive channel, when phonon converts into diffusive mode. 
	Throughout this paper we use Keldysh diagrammatic technique.\cite{keldysh,rammer}}
	\label{fig:sigma}
\end{figure}

We consider a dirty conductor with an electron action
\be
	\label{eq:action_n}
	S_{el,n}
	=\int dt(d\vec r)
	\psi_i^*\lb
	i\p_t
	-\lp
	\xi(\vec{p})+U(\vec r)
	\rp
	\rb
	\psi_i
\ee
with $U$ being disorder potential and $i=\uparrow,\downarrow$ corresponding to spin indices. Electron-phonon interaction is given by
\begin{align}
	\label{3}
	S_{el-ph,n}
	=\int dt(d\vec r)
	\psi_i^*
	\lb
	(p_Fv_F/d)\partial_\alpha u_\alpha(\vec r)-\right.
	\\
	\nn
	\left.
	-\partial_\alpha(u_\alpha(\vec r) U(\vec r))
	\rb
	\psi_i,
\end{align}
where the first term describes interaction due to the modulations of ionic density while the second term corresponds to distortions of disorder potential, the latter being described by a correlator
\be
	\label{disorder_u}
	\Braket{U(\vec{r})U(\vec{r}^\prime)}=u\delta(\vec{r}-\vec{r}^\prime).
\ee
It can be shown, using  comoving frame of reference \cite{blount,tsuneto,schmid,us,appendix}, 
that the action in the form of Eq.(\ref{3}) is equivalent to 
\begin{align}
	S_{el-ph,n}
	\simeq\int dt(d\vec r)
	\psi_i^*
	\Gamma_{n}^{\alpha\beta}
	(\partial_\beta u_\alpha)
	\psi_i,
\end{align}
with the vertex
\be
\label{eq:vertex_n}
	\Gamma^{\alpha\beta}_{n}=\left[p_\alpha v_\beta -(p_Fv_F/d)\delta_{\alpha\beta}\right],
\ee
where $p_F,\,v_F$ are Fermi momentum and velocity, $d$ is the dimensionality of the electron system.

\section{Ultrasound attenuation in  normal conductors.}

\emph{Local processes}
Attenuation rate $\alpha(\omega)$ is related with the imaginary part of the phonon self energy,
\be
	\alpha(\omega)=\frac{1}{\rho_m\omega}\Im\Sigma^R(\omega,q)\Big|_{\omega=sq},
	\label{def}
\ee
Local contribution to ultrasonic attenuation is due to processes which involve creation and annihilation of electron and phonon states on the spatial scale comparable to the elastic mean free path. Such a process is depicted by the diagram in Fig.2 a) and leads to a well known \cite{pippard,akhiezer,blount,tsuneto,schmid} result (it can be  obtained via the calculation of the diagram  Fig.~\ref{fig:sigma}a with  the vertices defined in Eq.~\ref{eq:vertex_n}):
\be
	\alpha_{n,l}=2c_l\frac{\nu p_F^2}{\rho_m}Dq^2
	\propto\omega^2,
	\label{alpha_l}
\ee
where $\nu$ is electron density of states per one spin, $D$ is diffusion coefficient and $c_l=2(d-1)/d(d+2)$ is just a numerical coefficient.

\emph{Energy diffusion processes.} 
For local processes information about direction of electron motion is retained between the acts of 
absorption and emission of phonons, thus the averaging of the product of two vertices 
over directions of momentum in the diagram~(Fig.~\ref{fig:sigma}a) leads to a large result:
\be
	\Big\langle\Gamma_n(\vec p)\Gamma_n(\vec p)\Big\rangle_{\vec p}\sim p_F^2v_F^2.
	\label{Gamma1}
\ee

However, when diffusion processes are allowed, the same information is lost between absorption and emission of phonons. On a formal level, it is seen from  the diagram (Fig.~\ref{fig:sigma}b) where an impurity ladder is inserted into the diagram disconnecting the electron-phonon vertices. Multiple collisions with impurities hold electrons in the regions of phonon absorption and emission leading to independent averaging of the interaction vertices $\Gamma$ over directions of momenta. That results in a much smaller values of effective vertex, 
which is now of the order of the temperature $T\ll p_Fv_F$,
\be
	\Big\langle\Gamma_{n}(\vec p)\Big\rangle_{\vec p}\sim \varepsilon\sim T
\ee
This is the reason  why diffusion modes are usually neglected in electron-phonon interaction in disordered conductors.

\begin{figure}[t]
	\center{\includegraphics[width=0.8\linewidth]{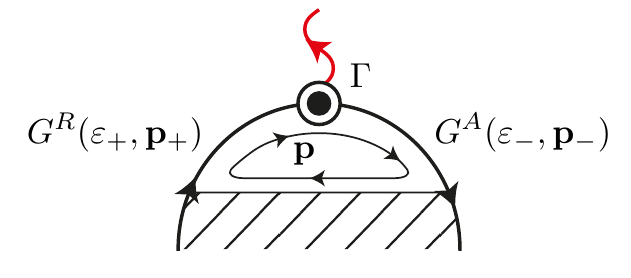}} 
	\caption{(Color online) When coupled to diffusive mode, electron vertex is always averaged over the momentum $\vec p$ that goes in the block of two Green function preceding the diffuson.}
	\label{fig1}
\end{figure}

It is convenient to define effective electron-diffuson vertex that depends on electron energy instead of a quickly relaxing momentum. Diagrammatically (see Fig.~\ref{fig:sigma}b) there is always a block of two Green functions that stands in between the phonon and diffusive mode and results in averaging over the electron momentum. Therefore an effective phonon-diffuson vertex $\Braket{\Gamma}_{n}$ may be defined with the help of the following integral representation:
\begin{align}
	\label{eq:gamma_introduced}
	\Braket{\Gamma}_{n}^{\alpha\beta}(\varepsilon)
	=
	\frac{1}{2\pi\nu\tau}\int(dp) G^R_-\lb\Gamma^{\alpha\beta}_{n}(\vec p_-,\vec p_+)\rb G^A_+,
\end{align}
where subscripts $\pm$ stand for $(\varepsilon\pm\omega/2, \vec{p}\pm\vec{q}/2)$ respectively.
In  general, effective vertex depends on both energies $\varepsilon_\pm=\varepsilon\pm\omega/2$ but in the limit \hbox{$\omega\ll\varepsilon\sim T$}) dependence on $\omega$ is negligible. 
New vertex $\Gamma_{ph-d,n}(\varepsilon_+,\varepsilon_-)$ is then given by
\begin{align}
	\label{eq:vertex_n_e}
	\Braket{\Gamma}^{\alpha\beta}_{n}(\varepsilon)
	=
	\varkappa \varepsilon\,\delta_{\alpha\beta},
\end{align}
where
\be
	\varkappa
	=
	\frac{\partial (pv/d)}{\partial\varepsilon}
	\equiv
	\lp
	1-\frac{p_Fv_F}{d}\frac{\partial\ln\nu}{\partial\varepsilon}
	\rp.
\ee

Diffusive electron modes evolve on a much larger timescales $\sim (Dq^2)^{-1}$ implying a more effective dissipation, despite  weak phonon-diffuson conversion. While the square of the vertex is smaller by the factor $(T/\varepsilon_F)^2$, the dissipation due to diffusion processes is enhanced by a factor $(ql)^{-2}$ (at the lowest frequencies enhancement is saturated by a factor $v_F^2/s^2$, $s$ being a sound velocity).

To the best of our knowledge, such processes for a single-band conductor under the condition of the perfect Coulomb screening had always been neglected so far. The corresponding  diagram is shown in  Fig.~\ref{fig:sigma}b. Its evaluation leads to the following result (see also Eq.(\ref{def})), valid at $\hbar\omega \ll T$: 
\be
	\label{eq:alpha_nd1}
	\alpha_{n,d}=\int_0^{\infty} \lp d\varepsilon\frac{\partial f(\varepsilon,T)}{\partial\varepsilon}
	\rp
	\alpha_{n,d}^{(\varepsilon)}(\omega),
\ee
where $\alpha_{n,d}^{(\varepsilon)}(\omega)$ is the partial contribution of electrons with energies in the interval $(\varepsilon,\varepsilon+d\varepsilon)$,   
\be
	\label{eq:alpha_nd2}
	\alpha_{n,d}^{(\varepsilon)}(\omega)=\frac{q^2}{\rho_m}
	\left(\varkappa\varepsilon\right)^2\lp2\nu\Re\D(\omega,q)\rp.
\ee
Eq.(\ref{eq:alpha_nd2}) contains   diffusion propagator $\D(\omega,q)$ (the corresponding diagrams are shown in Fig.~\ref{fig:diff}) equal to
\be
	\D(\omega,q)=\frac{1}{-i\omega+Dq^2},
\ee
where $\tau=l/v_F$ is an electron elastic scattering time and the diffusion coefficient $D=\tau v_F^2/d$. (Note that $\tau=(2\pi\nu u)^{-1}$ with electron density of states $\nu$ and $u$ defined in Eq.\ref{disorder_u}) Calculation of the integral in Eq.(\ref{eq:alpha_nd1}) leads to the following result for the ultrasound attenuation rate at frequency $\omega$ and temperature $T$:
\begin{align}
	\label{alpha_d2}
	\alpha_{n,d}
	=
	\frac{2\pi^2}{3}\frac{\nu D}{\rho_ms^4}
	\frac{\varkappa^2\,T^2\omega^2}{1 + (D \omega/s^2)^2}.
\end{align}
It is useful to present it also in the form of the ratio of (\ref{alpha_d2}) to the "local" result, Eq. (\ref{alpha_l}):
\be
	\frac{\alpha_{n,d}}{\alpha_{n,l}}=
	\frac{\pi^2}{3c_l}
	\lp\frac{ \varkappa T}{p_F s}\rp^2\frac{\omega_c^2}{\omega_c^2 + \omega^2}\,; \qquad \quad\omega_c = \frac{s^2}{D}
	\label{ratio1}
\ee
The two  mechanisms of dissipation  lead also to  different dependencies of attenuation on temperature and frequency:  "local" attenuation rate is only weakly $T$-dependent but grows as $\omega^2$; On the other  hand, attenuation due to "diffusive"  mechanism is \hbox{proportional to $T^2$}. Dissipation due to diffusive mechanism is frequency-indepedent at high $\omega \gg \omega_c = s^2/D$ and goes as $\omega^2$ at low frequencies. Formulae (\ref{alpha_d2},\ref{ratio1}) can be used in order to extract the value of the electron diffusion constant $D$ from the data on the ultrasound attenuation.

For thermal phonons with $\hbar \omega \sim T$ attenuation due to diffusive channel is always small, $\alpha_{n,d}\ll\alpha_{n,l}$. However, for the ultrasonic attenuation at low frequencies \, $\omega \ll T/\hbar$ the situation can be quite different, especially in  doped semiconductors at moderate temperatures. Consider, for example, heavily doped Si with $n=10^{20}cm^{-3}$, $m=0.36m_0$, $p_Fl=10$, $s\approx8\cdot10^5cm/s$. At the temperature $T=0.1 E_F \approx200K$ one finds
\be
	\frac{\alpha_{n,d}}{\alpha_{n,l}}\approx100,\quad
	f=2\pi\omega\leq 10 GHz
\ee
i.e. attenuation is enhanced by two orders of magnitude due to contribution of the diffusive channel.

\section{S-wave superconductors} 

\subsection{Formulation of the model}

We consider BCS superconductor (SC) with an s-wave pairing and microscopic interaction constant  $g$,
\be
	S_{el,sc-int}=\int dt(d\vec r)\lb
	g\psi^*(\vec r)\psi^*(\vec r)\psi(\vec r)\psi(\vec r)
	\rb.
\ee

\begin{figure}[t]
	\center{\includegraphics[width=1\linewidth]{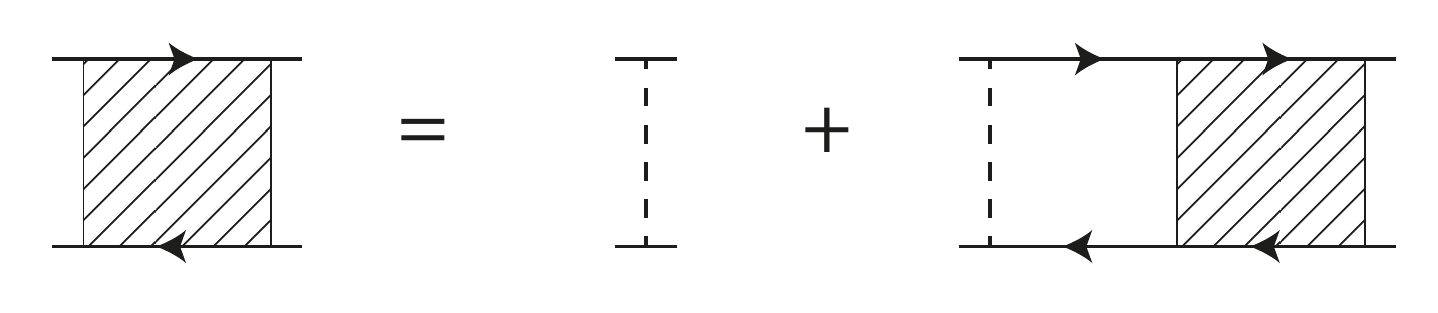}} 
	\caption{(Color online) Diffusive modes correspond to impurity ladders. The quantity $\D(\dots)$ in the text is given by the sum of this ladder, excluding one overall $(2\pi\nu\tau)^{-1}$ factor coming from the first impurity line.}
	\label{fig:diff}
\end{figure}

In the case of a dirty superconductor it is convenient to define Green function as 
\begin{align}
	\label{G_swave}
	\check{G}^R_{\alpha\beta}(\varepsilon,\vec p)=&-i\Braket{\Psi_\alpha\overline{\Psi}_\beta}
	\\
	\nn
	=&
	[(\varepsilon+i0)\check{\tau}_3-\xi\check{\tau}_0-\Delta (i\check{\tau}_2) ]_{\alpha\beta}^{-1},
\end{align}
where 4-component spinor $\Psi$ is defined as follows:
\begin{align}
	\label{psi_spinors}
	\Psi(\vec p)
	&=
	\frac{1}{\sqrt{2}}
	\begin{pmatrix}
	\psi_{\vec p \uparrow}
	\\
	\psi^*_{-\vec p \downarrow}
	\\
	\psi_{\vec p \downarrow}
	\\
	-\psi^*_{-\vec p \uparrow}
	\end{pmatrix}
	,\,\quad
	\overline{\Psi}=\Psi^+\check{\tau}_3.
\end{align}
The 4-dimensional Nambu-Gor'kov space corresponds to the product of spin and particle-hole spaces. However, for our purposes only particle-hole subspace is relevant, as is evident from Eq.(\ref{G_swave}), where only particle-hole Pauli matrices $\check{\tau}$ are present. We will thus ignore the spin structure, effectively working with 2-component $\Psi,\overline{\Psi}$ spinors.

For a disordered s-wave superconductor self-consistent Born approximation gives\cite{AGD}
\be
	\check{G}^R(\varepsilon,\vec p)
	=
	\lb\eta(\varepsilon)\Big(\varepsilon\check{\tau}_3-\Delta(i\check{\tau}_2)\Big)-\xi\check{\tau}_0 \rb^{-1}
	\label{G-sc}
\ee
with
\begin{align}
	\eta(\varepsilon)=\lp1+\frac{i}{2\tau E}\rp,\quad
	E=\sqrt{\varepsilon^2-\Delta^2}.
\end{align}

\subsection{Energy density mode}

\subsubsection{Diffuson}

Diffusion modes are described by impurity ladder shown on the Fig.(\ref{fig:diff}). This ladder can be analyzed with the help of a Bethe-Salpeter equation 
\be
	(\underline{\hat{\D}})^{-1} (\varepsilon,\omega,q)=\underline{\hat{1}}-\underline{\hat{\Xi}}(\varepsilon,\omega,q)
	\label{Dmat}
\ee
with the self-energy
\be
	\underline{\hat{\Xi}}(\varepsilon,\omega,q)=u\int(d\vec p)\,\check{G}^R(\varepsilon_-,\vec p)\otimes \check{G}^A(\varepsilon_+,\vec p).
	\label{Sigma1}
\ee
Green function is a $2\times2$ matrix in the particle-hole space (2-dimensional spin space is irrelevant, \hbox{see Eq.(\ref{psi_spinors})}). Thus, $\underline{\hat{\D}}$ and $\underline{\hat{\Xi}}$ are matrices in a $4$-dimensional space constructed as a product of two $2$-dimensional particle-hole subspaces.
These  $4\times4$ matrices $\underline{\hat{D}}$ and $\underline{\hat{\Xi}}$ could be interpreted as superoperators acting on $2\times2$ operators such as effective electron-phonon vertex $\check{\Gamma}$ (for an s-wave state $\check{\Gamma}$ is defined further in this chapter, see Eq.(\ref{effective_gamma_s})). 
For \hbox{example}, if $\check{A}$, $\check{B}$ and $\check{X}$ are $2\times2$ matrices, then $\check{Y}=\check{A}\check{X}\check{B}$ is as well a matrix of this type. This means that $\check{A}\otimes\check{B}$ describes a mapping $\check{X}\rightarrow\check{Y}$ thus indeed being a superoperator.

At zero external frequency and momentum the self-energy reads as follows:
\begin{align}
	\underline{\hat{\Xi}}=\frac{1}{2}\check{\tau}_0\otimes\check{\tau}_0
	+
	\frac{
		(\varepsilon\check{\tau}_3-\Delta(i\check{\tau}_2))\otimes(\varepsilon\check{\tau}_3-\Delta(i\check{\tau}_2))
	}
	{2(\varepsilon^2-\Delta^2)}.
\end{align}
Analysing this matrix structure one finds  two massless modes corresponding to operators
\be
	\check{\tau}_0
	\quad\text{and}\quad
	\varepsilon\check{\tau}_3-\Delta(i\check{\tau}_2).
\ee
The $\check{\tau}_0$ mode is symmetric in particle-hole space and corresponds to charge density, while the asymmetric mode~$\varepsilon\check{\tau}_3+\Delta(i\check{\tau}_2)$ corresponds to energy density. The corresponding term in the matrix propagator defined in  Eq.(\ref{Dmat})  is given by
\be
	(\underline{\hat{\D}}) (\varepsilon,\omega,q)
	\rightarrow
	\frac{1}{2\pi\nu\tau^2}\D_s^{(\varepsilon)}(\omega,q),
\ee
where 
\begin{align}
	\label{D_swave}
	\D_s^{(\varepsilon)}(\omega,q)=&\frac{1}{\lb-i(E_+-E_-)+Dq^2\rb}
\end{align}
and $E_\pm= \sqrt{(\varepsilon \pm \omega/2)^2-\Delta^2}$. Charge density fluctuations are prohibited by strong Coulomb interaction, so we should not consider them here.

Strictly speaking,  there are two more diffusion modes of the  "Cooperon" type which are related to the charge conversion
between condensate and thermal excitations. Contrary to  Eq.(\ref{Sigma1}), these modes
are constructed with matrix products of the type of $\check{G}^R(\varepsilon_-,\vec p)\otimes \check{G}^A(-\varepsilon_+,\vec p)$. The corresponding propagator is proportional to \hbox{$(-i(E_++E_-)+Dq^2)^{-1}$}. It is important that in the limit $Dq^2\ll\sqrt{T\Delta}$ the contribution from these modes is much smaller than that of the energy diffusion channel, so we neglect Cooperon channel in the following.

\subsubsection{Effective vertex}

When defining equivalent phonon-diffuson vertex we now have to pay attention to the matrix structure in the Nambu-Gor'kov space ($\check{\Gamma}_n=\Gamma_n\cdot\check{1}$),
\begin{align}
	\label{effective_gamma_s}
	\Braket{\check{\Gamma}}_{s}(\varepsilon)
	=
	\int(dp)\check{G}^R_-\Gamma_{n}(\vec p_-,\vec{p}_+)\check{G}^A_+,
\end{align}
that modifies the vertex (\ref{eq:vertex_n_e}) into
\begin{align}
	\Braket{\check{\Gamma}}_{s}(\varepsilon)
	=
	\varkappa
	\Big(
	\varepsilon\check{\tau}_3-\Delta(i\check{\tau}_2)
	\Big)\delta_{\alpha\beta}.
	\label{SCvert}
\end{align}
We see that there is indeed no coupling between ultrasound and charge density mode.

Generally, acoustic wave modifies effective BCS coupling constant $\lambda$ leading to an additional electron-phonon vertex. This vertex has the same  matrix structure as the order parameter:
\be
	\label{eq:lambda}
	\check{\Lambda}=\varkappa_\Delta(\Delta\check{\tau}_1)\delta_{\alpha\beta}
\ee
with constant $\varkappa_\Delta$ being equal to
\be
	\label{eq:vertex_lambda2}
	\varkappa_\Delta=-
	\frac{d \Delta}{d\ln\rho}
	\,\underset{BCS}{=}\,
	-\frac{1}{\lambda}\left(
	\frac{d \ln\lambda}{d\ln\rho}\right).
\ee
Here $\lambda=\nu g$ is a dimensionless BCS coupling constant. Below in this section we consider temperatures and all other relevant energy scales to be well below the gap $T,\,\omega,\,Dq^2\ll\Delta$.

Variations of $\lambda$ arise either directly through the density of states $\nu$ or through interaction constant $g$, 
\be
	\frac{d \ln{\lambda}}{d\ln\rho}
	=
	\frac{d \ln{\nu}}{d\ln\rho}+
	\frac{d \ln{g}}{d\ln\rho}.
\ee
Within our model $d\ln\nu/d\ln\rho=(p_Fv_F/d)(d\ln\nu/dE)$; it  arises due to the shift of the chemical potential in the presence of acoustic wave.

Defining effective vertex $\Braket{\check{\Lambda}}_{s}$ analogously to Eq.(\ref{effective_gamma_s}) we get
\be
	\Braket{\check{\Lambda}}_{s}(\varepsilon)
	=
	\varkappa_\Delta\lp\frac{\Delta^2}{\varepsilon^2-\Delta^2}\rp
	\Big(
	\varepsilon\check{\tau}_3-\Delta(i\check{\tau}_2)
	\Big)\delta_{\alpha\beta}.
\ee
The effective vertex $\Braket{\check{\Lambda}}_{s}$ is substantially enhanced due to singular density of states, in contrast with the vertex $\Braket{\check{\Gamma}}_{s}$.

\subsection{Ultrasound attenuation  due to energy diffusion}

Similarly to the normal metal case,  contribution of diffusion channel to ultrasonic attenuation is given by the following integral (assuming $\omega \ll T$):
\be
	\label{alpha_sd}
	\alpha_{s,d} (\omega) =\int\limits_0^{\infty} \lp d\varepsilon\frac{\partial f(\varepsilon,T)}{\partial\varepsilon}
	\rp
	\alpha_{s,d}^{(\varepsilon)}(\omega)
\ee
where $\alpha_{s,d}^{(\varepsilon)}(\omega)$ is the partial contribution of quasiparticles with energies in the range ($\varepsilon,\varepsilon+d\varepsilon$),
\be
	\label{alpha_sd-e}
	\alpha_{s,d}^{(\varepsilon)} (\omega) =\frac{q2}{\rho_m}
	\lp
	\varkappa E+\varkappa_\Delta \frac{\Delta^2}{E}
	\rp^2
	\lp2\nu_n\Re\D_s^{(\varepsilon)}(\omega,q)\rp,
\ee
$\D_s^{(\varepsilon)}(\omega, q)$ being a diffuson in a superconducting state given by Eq.(\ref{D_swave}) and $E=\sqrt{\varepsilon^2-\Delta^2}$. We have also added subscript (n) to normal metal DOS $\nu_n$ to avoid possible confusion. Note that the contribution due to variations of the  effective BCS interaction constant  (described by the vertex  $\Braket{\check{\Lambda}}$) is enhanced due to singularity in the density of states at the gap edge,
\mbox{$\nu(\varepsilon)/\nu_n\simeq\Delta/\sqrt{\varepsilon^2-\Delta^2}$}.

Substitution of Eq.(\ref{alpha_sd-e}) into Eq.(\ref{alpha_sd}) and  integration yields final result for ultrasonic attenuation due to energy diffusion channel in s-wave superconductors:
\begin{widetext}
	\begin{equation}
		\label{alpha_s2}
		\alpha_{s,d} (\omega)
		=
		2\frac{\nu }{\rho_mD}
		\exp\lb-\frac{\Delta}{T}\rb
		\begin{cases}
		(\Delta/T)\left(
		\varkappa_\Delta^2\Delta^2\ln\frac{T}{A(\omega)}
		+4\varkappa\varkappa_\Delta\Delta T
		+4\varkappa^2 T^2
		\right) & \omega\gg \omega_c\sqrt{\dfrac{\Delta}{T}}
		\\
		2\left(
		\varkappa_\Delta^2\Delta^2
		+4\varkappa\varkappa_\Delta\Delta T
		+8\varkappa^2 T^2
		\right)
		\times
		\lp
		\dfrac{D\omega}{s^2}
		\rp^2
		&
		\omega\ll  \omega_c\sqrt{\dfrac{\Delta}{T}}
		\end{cases},
	\end{equation}
\end{widetext}
where $A(\omega) \equiv  \omega + \Delta (\omega_c/\omega)^2$ and $\omega_c=s^2/D$ is the crossover frequency in the normal state.
Equations (\ref{alpha_s2}) were derived under the conditions
\be
	\omega,Dq^2\ll T\ll\Delta\ll\tau^{-1}\ll p_Fv_F
\ee
and contain two characteristic crossover frequencies, $\omega_{c,s1} = \omega_c(\Delta/T)^{1/2}$ , and $\omega_{c,s2} = \omega_c^{2/3}\Delta^{1/3}$.

Note that $\omega_{c,s2}$ describes a weak logarithmic crossover that results from the DOS singularity in the superconducting state; $(\Delta/T)$ factor in the first line of Eq.(\ref{alpha_s2}) is the result of DOS behaviour as well. This crossover exists only if $\omega_{c,s2}>\omega_{c,s1}$ or equivalently only for temperatures $T>(\Delta\omega_c^2)^{1/3}$. The frequency of the second crossover $\omega_{s,c2}$ can be small or large in comparison with
$\omega_{s,c1}$, depending on specific material.

To compare attenuation due to the energy diffusion channel (\ref{alpha_s2})  with  the one produced by usual local processes in superconducting state (we denote it as $\alpha_{s,l}$) note that the latter is proportional to the density of normal electron-hole excitations~\cite{tinkham}. Thus
the ratio 
\begin{equation}
	\label{alpha_sl}
	\frac{\alpha_{s,l}}{\alpha_{n,l}}= 
	\int_\Delta^\infty d\varepsilon\,\frac{\partial f(\varepsilon)}{\partial\varepsilon}
	\simeq2\exp\lb-\frac{\Delta}{T}\rb,
\end{equation}
so that $\alpha_{s,l}\propto\omega^2$ as in the normal state. At high frequencies $\omega \gg \omega_{c,s2}$  we thus find that the role of energy diffusion channel grows with $\omega$ decrease:
\begin{align}
	\frac{\alpha_{s,d}}{\alpha_{s,l}}
	=
	\frac{1}{2c_l}\frac{1}{p_F^2s^2}
	\Big(&
	\varkappa_\Delta^2\frac{\Delta^3}{T}\ln\frac{T}{A(\omega)}
	+4\varkappa\varkappa_\Delta\Delta^2+
	\\
	\nn
	&+4\varkappa^2\Delta T
	\Big)
	\times
	\lp\dfrac{\omega_c}{\omega}\rp^2.
\end{align}
In this frequency range attenuation due to energy diffusion  has only weak logarithmic frequency dependence; it also depends on temperature in nontrivial way that does not reduce to the $T$-dependence of normal electron density.

At lowest frequencies, when $\omega \leq \omega_{c,s1}$, frequency dependence of $\alpha_s$ is of standard form, but temperature dependence differs from that of local contribution:
\begin{align}
	\frac{\alpha_{s,d}}{\alpha_{s,l}}
	=
	\frac{1}{c_l}\frac{1}{p_F^2s^2}
	\Big(&
	\varkappa_\Delta^2\Delta^2
	+2\varkappa\varkappa_\Delta\Delta T+8\varkappa^2 T^2
	\Big).
	\label{rs}
\end{align}

Inspection of the ratio (\ref{rs}) shows that usually it is rather small; interesting exception is presented by superconductors with extremely low electron density, which are not far from the crossover to the regime of "local pairs". Particular example of this kind is presented by recently discovered heavy-metal compound YPtBi \cite{yptbi} with conduction electron density at temperatures about Kelvin range as low as $n = 2\cdot10^{18} cm^{-3}$. With the values of other parameters taken from Ref.~\cite{yptbi} as $m=0.15m_0$, electron mean free path $l=130nm$,  $T_c=0.77K$, $\omega_D=200K$ and sound velocity\cite{yptbi_s} $s=2\cdot10^5cm/s$, we find for $T=0.2K$ the ratio of the energy diffusion contribution to the standard PIC result $\alpha_{s,l}$:
\begin{align}
\frac{\alpha_{s,d}}{\alpha_{n,l}}\approx1,\quad f=2\pi\omega\le 1GHz.
\end{align}

\section{D-wave superconductors}

\subsection{Hamiltonian}

\begin{figure}[t]
	\center{\includegraphics[width=1\linewidth]{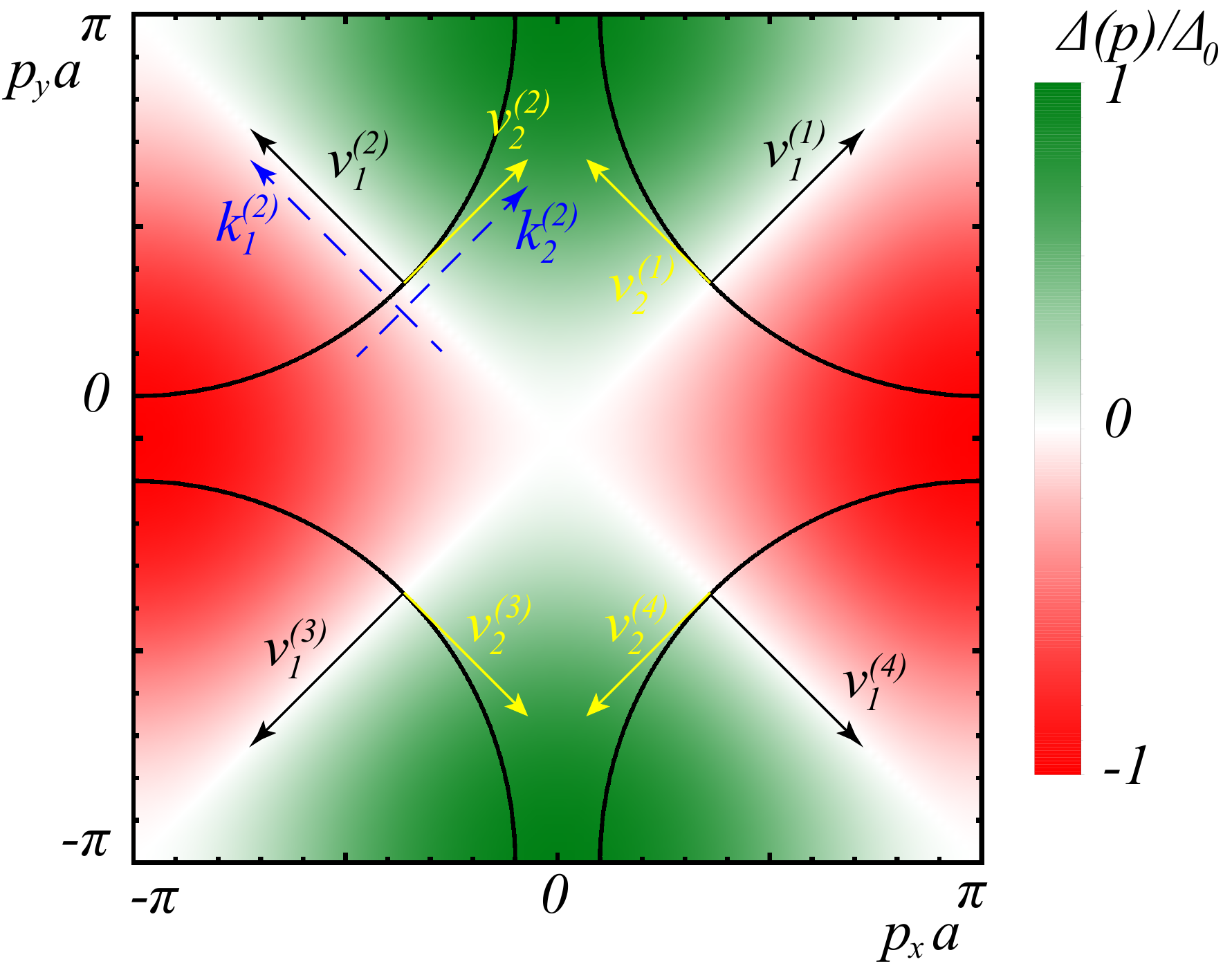}} 
	\caption{(Color online) Schematic Fermi surface of a d-wave superconductor. Background color represents order parameter $\Delta(\vec p)$ angular dependence revealing four nodes on the Fermi surface. Each of them has its own electron Fermi $\vec{v}_1=d\xi/d\vec{p}$ and gap $\vec{v}_2=d\Delta/d\vec{p}$ velocities. We thus choose four different coordinate systems $(k_1,k_2)^{(i)}$(for clarity only one is shown) around each node with axes aligned with local electron Fermi and gap velocities. $x,y$ directions correspond to [100], [010] crystalline axes; $a$ is the lattice constant.}
	\label{fig:dwave_FS}
\end{figure}

\begin{figure}[b]
	\center{\includegraphics[width=1\linewidth]{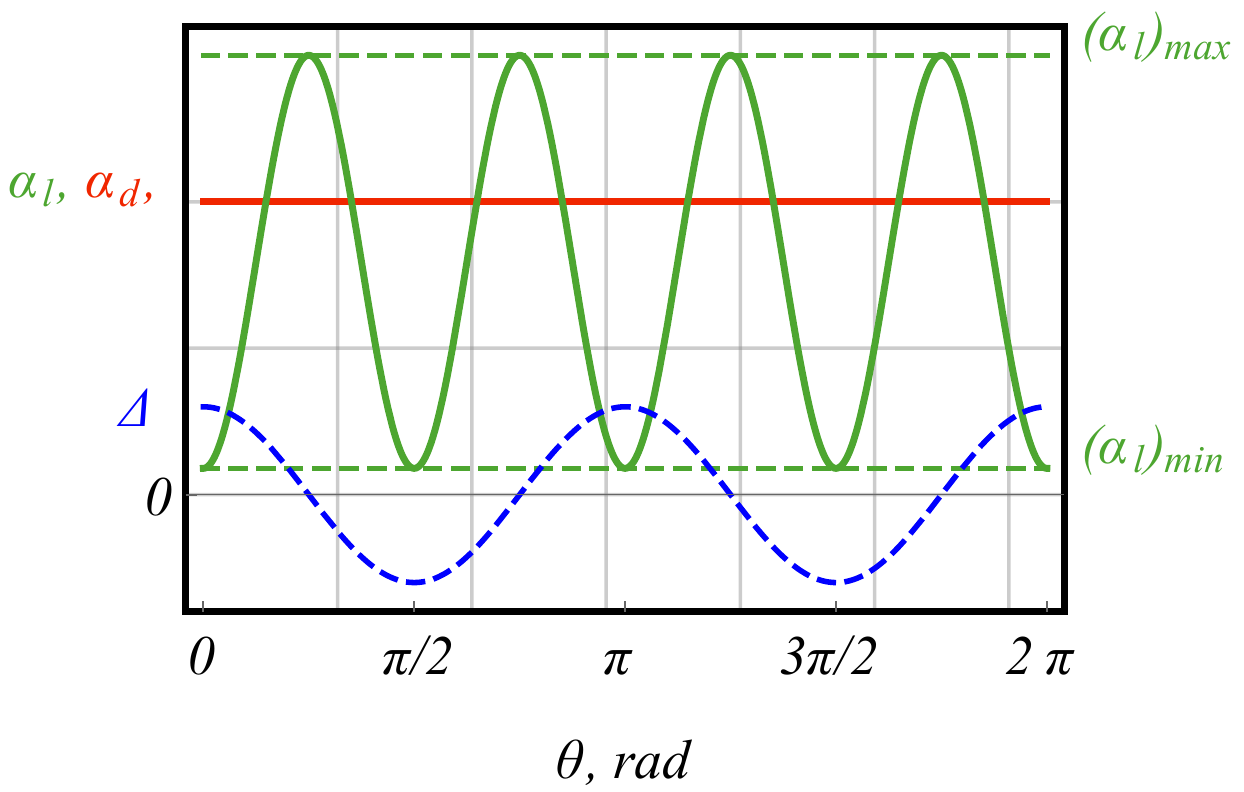}} 
	\caption{(Color online) Schematic angular dependence of ultrasonic attenuation for local ($\alpha_l$) and diffusive ($\alpha_d$) processes. $\alpha_l$ exhibits strong anisotropy with extremal values $(\alpha_l)_{min},(\alpha_l)_{max}$ shown by horizontal dashed lines. In contrast to local processes, attenuation in diffusion channel is isotropic. \hbox{As a reference} direction $\theta=0$ we choose [110]. Minima of ultrasonic attenuation are observed in $\pm$[100] and $\pm$[010] directions, \hbox{while maxima in $\pm$[110] and $\pm$[1-10]}. The angular dependence of the gap $\Delta(\vec k)$ is shown as well (dashed blue line).}
	\label{fig:dwave_angular}
\end{figure}

We consider here strongly a model of anisotropic d-wave superconductor and neglect electron dispersion in the direction
transverse to the layers. Then the order parameter depends on  direction of momentum(Fig.\ref{fig:dwave_FS})  in the following way:
\be
	\Delta(\vec p)=\Delta_0(\cos p_xa-\cos p_ya),
\ee
where $a$ is the lattice constant. At low temperatures $T\ll\Delta_0$
this leads to angular dependence of ultrasonic attenuation due to local processes. At the same time, the contribution of diffusion channel is isotropic (see Fig.(\ref{fig:dwave_angular})).

For d-wave superconductor it is convenient to define Green functions somewhat differently from s-wave case, namely:
\begin{equation}
	\check{G}^R_{\alpha\beta}(\varepsilon,\vec p)=-i\Braket{\Psi_\alpha\Psi^+_\beta}
	=
	[(\varepsilon+i0)\check{\tau}_3-\xi-(i\check{\tau}_2)\Delta ]^{-1},
	\label{G-scd}
\end{equation}
see the difference with Eq.(\ref{G-sc}), where we had used  $\overline{\Psi}\equiv\check{\tau}_3\Psi^+$ instead of $\Psi^+$.
Energy  of excitations above the d-wave SC state  vanishes at 4 nodal directions in the momentum space of the order parameter, see Fig.(\ref{fig:dwave_angular}).  Near these directions the excitation spectrum can be linearized:
\be
	\xi=v_Fk_1,\,\Delta=v_gk_2,
\ee
where momentum $\vec{k}$ is measured from the node, $\vec{k}=\vec{p}-\vec{p}_{node}$, the basis is individual for each node (see the  Fig.\ref{fig:dwave_FS}).   Due to the presence of low-lying excitations, all thermal effects at $T \ll T_c$ are described by some power laws of $T$, instead of an exponential behavior in s-wave case.
Now we should take into account random disorder potential, which we consider in the white-noise limit:
\be
	\Braket{U(\vec{r})U(\vec{r^\prime})}=u\delta(\vec{r}-\vec{r}^\prime)
\ee
Scattering by disorder results in renormalization of Green functions, leading both to a scattering rate $\gamma$ and to renormalization of the energy:  $\varepsilon\rightarrow\tilde{\varepsilon}$. In the self-consistent Born approximation we find
\begin{align}
	\Braket{\check{G}^R_{\alpha\beta}(\varepsilon,\vec p)}
	=
	[(\tilde{\varepsilon}+i\gamma)\check{\tau}_3-\xi\check{\tau}_0+(i\check{\tau}_2)\Delta ]^{-1},
\end{align}
where $\tilde{\varepsilon}$ is related  to a residue of the Green function,
\be
	\varepsilon\equiv Z(\varepsilon)\tilde{\varepsilon}(\varepsilon).
\ee
Quantities $\gamma$ and $\tilde{\varepsilon}$ are to be determined by  self-consistency equations \cite{koshelev}
\begin{align}
	\label{dwave_gamma}
	1&=K
	\left(
	\frac{\tilde{\varepsilon}}{\gamma}\arctan\frac{\tilde{\varepsilon}}{\gamma}+\ln\frac{\Delta_0}{\sqrt{\tilde{\varepsilon}^2+\gamma^2}}
	\right),
	\\
	\label{dwave_etilde}
	\varepsilon&=K\frac{\tilde{\varepsilon}^2+\gamma^2}{\gamma}\arctan\frac{\tilde{\varepsilon}}{\gamma},
\end{align}
where $K$ is a dimensionless disorder strength,
\be
\label{K}
	K=\frac{Nu}{2\pi v_Fv_g}\equiv N\frac{v_F}{v_g}G_n^{-1}.
\ee
Here $N$ is a number of valleys ($N=4$), and $G_n$ is the dimensionless conductance per layer in the normal state (in units of $e^2/\hbar$). Self-consistent Born approximation is valid as long as $K\ll1$; since the ratio $v_F/v_g$ is a large parameter (according to Ref.~\cite{bscco_vratio}, it is about 20 for BSCCO superconductors), thus normal-state conductance should be sufficiently large, $G_n \geq 100$, for this approach to be valid.

Asymptotic behaviour of $\gamma(\varepsilon)$ and $Z(\varepsilon)$ is of particular interest, 
it is given by 
\begin{align}
	\label{dwave_asymptotics_low}
	\varepsilon\ll\gamma_0:\, & \gamma(0)=\Delta_0e^{-1/K}, & Z(0)=K,
	\\
	\label{dwave_asymptotics_high}
	\varepsilon\gg\gamma_0:\, & \gamma=\dfrac{\pi}{2K}\dfrac{\varepsilon}{(\ln(\varepsilon/\gamma_0))^2}, & Z(\varepsilon)=K\ln(\varepsilon/\gamma_0),
\end{align}
where $\gamma_0\equiv\gamma(0)$ is a scattering rate at the Fermi surface. Similar results were obtained previously for graphene~\cite{ostrovsky}. Below we restrain ourselves to the intermediate temperature range $\gamma_0 \ll T \ll \Delta_0$.

To describe  electron-phonon interaction, it is still sufficient\cite{appendix} to start from the electron stress tensor Eq.(\ref{eq:vertex_n}), 
but now a momentum-dependence of the order parameter $\Delta(\vec p)$ should be taken into account:
\begin{align}
	\label{vertex_dwave}
	&\check{\Gamma}_{d}^{\alpha\beta}=p_\alpha\partial_\beta\check{H}-\frac{p_Fv_F}{2}\delta_{\alpha\beta}
	\\
	\nn
	&=
	\begin{pmatrix}
		(v_Fp_F/2)+v_Fk_F & (p_F+k_F)v_g(i\check{\tau}_2)
		\\
		k_gv_F & -(p_Fv_F/2)+k_gv_g(i\check{\tau}_2)
	\end{pmatrix}_{\alpha\beta},
\end{align}
where we have used $\check{H}=\xi\check{\tau}_0+\Delta(i\check{\tau}_2)$.

\subsection{Ultrasound attenuation due to local processes.}

Like previously, attenuation due to local processes in a d-wave superconductor can be depicted by the electron bubble diagram (Fig.\ref{fig:sigma}a), the corresponding analytical expression reads as
\be
	\alpha_{d,l}=\frac{q^2}{2\rho_m\omega}
	\Tr\lb(f_+-f_-)\check{\Gamma}(\check{G}^R_--\check{G}^A_-)\check{\Gamma}(\check{G}^R_+-\check{G}^A_+)\rb,
\ee
where $\check{\Gamma}$ is an abbreviation for
\be
	\check{\Gamma}=
	e_{\alpha}\check{\Gamma}^{\alpha\beta}_{d,l} e_{\beta}
\ee
with $e_\alpha$ being a unit vector in the direction of phonon momentum $q_\alpha$ (while polarization vector of longitudinal phonons is essentially the same as $e_{\alpha}$). 
In contrast to normal and s-wave states,  in d-wave state attenuation is essentially anisotropic due to two strongly different velocities $v_F,v_g$ and  symmetry of the order parameter.

Partial contribution of electrons with energies $(\varepsilon,\varepsilon+d\varepsilon)$ into attenuation rate can be found in the
form
\be
	\label{alpha_local_dwave}
	\alpha_{d,l}^{(\varepsilon)}(\omega) =
	\frac{q^2}{2\pi^2 v_Fv_g\rho_m}
	\lp
	1+\frac{Z(\varepsilon)}{K} 
	\rp
	\times\Big( p_F^2v_F^2F(\theta)\Big),
\ee
where the function $F(\theta)$ comes from electron-phonon vertices and describes angular dependence of attenuation,
\be
	F(\theta)=\sin^22\theta+\lp\frac{v_g}{v_F}\rp^2\cos^22\theta.
\ee

Integration of Eq.(\ref{alpha_local_dwave}) over electron energy (like in Eq.(\ref{eq:alpha_nd1})) eventually leads to
\be
	\label{alpha_dwave_local}
	\alpha_{d,l}=
	\frac{F(\theta)}{2\pi^2}\frac{v_F}{v_g}\frac{p_F^2\omega^2}{\rho_ms^2}
	\ln\frac{T}{\gamma_0}.
\ee

Comparing of the result \eqref{alpha_dwave_local}  with attenuation in the (isotropic) normal state we find
\be
	\label{alpha_dwave_to_normal}
	\frac{\alpha_{d,l}}{\alpha_{n,l}}=2KF(\theta)
	\ln\frac{T}{\gamma_0},
\ee
where $K$ is the dimensionless measure of disorder defined in Eq.(\ref{K}) and $\gamma_0$ is the scattering rate at the Fermi surface, Eq.(\ref{dwave_asymptotics_low}). Note that in the limit $T\ll\gamma_0$ the contribution from local processes is very similar
to  \eqref{alpha_dwave_local}, one should just replace $\ln(T/\gamma_0)\rightarrow2$.

Attenuation rate Eq.(\ref{alpha_dwave_local}) possesses unusual logarithmic temperature behaviour $\propto\ln(T/\gamma_0)$ that does not correspond to naive expectations for the linear density of states, $\alpha\propto T$.    The reason is that electron scattering rate $\gamma$ is intimately related to the electron density of states; typical scattering rate decreases with $T$ together with
typical DoS. Namely, Eq.(\ref{dwave_asymptotics_high}) shows that at moderately high temperatures $\gamma\propto T$ up to a logarithmic factor.

\subsection{Energy diffusion mode}

\subsubsection{Diffusion propagator.}

This section proceeds in a way almost identical to s-wave case.
The diffusion modes are described by a Bethe-Salpeter equation Fig.(\ref{fig:diff})
\be
	(\underline{\hat{D}})^{-1} (\varepsilon,\omega,q)=1-\underline{\hat{\Xi}}(\varepsilon_-,\varepsilon_+,\vec q),
\ee
with a self-energy
\be
	\underline{\hat{\Xi}}(\varepsilon,\omega,\vec q)=u\sum_{nodes}\int(d\vec p)\check{G}^R_-\otimes \check{G}^A_+.
\ee

Energy density diffusion mode corresponds to $\check{\tau}_3$ eigenvector and eigenvalue of the self-energy
\be
	\underline{\hat{\Xi}}(\varepsilon,\omega,\vec q)
	\rightarrow
	u\sum_{nodes}\int(d\vec p)\Tr\lb\check{\tau}_3\check{G}^R_-\check{\tau}_3\check{G}^A_+\rb.
\ee
Summation of the impurity ladder for the energy density mode thus gives
\be
	(\underline{\hat{\D}}) (\varepsilon,\omega,\vec{q})
	\rightarrow
	2u\gamma(\varepsilon)\D_d^{(\varepsilon)}(\omega,q),
\ee
where $\D_d^{(\varepsilon)}$ is a diffuson propagator in a d-wave state,
\be
	\D_{d}^{(\varepsilon)}(\omega,q)=\frac{1}{-i\omega+D_d^{(\varepsilon)}q^2}.
\ee
The diffusion coefficient is
\be
	\label{dwave_D}
	D_d^{(\varepsilon)}=\frac{K+Z(\varepsilon)}{2\gamma(\varepsilon)}\Braket{v^2},
\ee
where $\Braket{v^2}=(v_F^2+v_g^2)/2\simeq v_F^2/2$.

Cooperon modes in a d-wave state are irrelevant at temperatures $T\gg\gamma_0$.

\subsubsection{Effective vertex}

Effective phonon-diffuson vertex is defined in the way similar to the s-wave state:
\begin{align}
	\Braket{\check{\Gamma}}_{d}^{\alpha\beta}(\varepsilon_+,\varepsilon_-)
	=
	u\sum_{nodes}
	\int(dp)\check{G}^R_+\check{\Gamma}_{d}^{\alpha\beta}(\vec p)\check{G}^A_-.
\end{align}
Calculation of this vertex requires some caution, see Appendix for details. Eventually we get, similar to the normal metal state:
\be
	\check{\Gamma}_{d,d}^{\alpha\beta}(\varepsilon)=\varepsilon\check{\tau}_3\delta_{\alpha\beta}.
\ee
The effective vertex describing coupling to the energy density mode turns out to be isotropic $\propto\delta_{\alpha\beta}$ once the summation over 4 nodes is performed. This happens despite the anisotropy of velocities for each node separately.
\newline

\subsection{Ultrasound attenuation due to energy diffusion}

We obtain the contribution of diffusion processes in a complete analogy to the s-wave state:
\be
\label{alpha_dwave_diff-e}
\alpha_{d,d}(\omega)=\int_0^\infty \lp d\varepsilon\frac{\partial f(\varepsilon,T)}{\partial\varepsilon}
\rp
\alpha_{d,d}^{(\varepsilon)}(\omega)
\ee
where
\be
	\label{alpha_dwave_diffusion}
	\alpha_{d,d}^{(\varepsilon)} (\omega) =\frac{q^2}{\rho_m}\cdot
	\varepsilon^2\cdot\Big(2\nu_d(\varepsilon)\Re\D_d^{(\varepsilon)}(\omega,q)\Big),
\ee
where $\nu_d(\varepsilon)=\gamma(\varepsilon)/\pi u$ is a density of states in a d-wave superconductor. Then Eq.(\ref{alpha_dwave_diffusion}) gives
\begin{align}
	\alpha_{d,d}^{(\varepsilon)}(\omega)=\frac{1}{\pi^2}\frac{v_F}{ v_g}\lp1+\frac{Z(\varepsilon)}{K}\rp\frac{\varepsilon^2}{\rho_ms^4}
	\frac{
		\omega^2}{1+\omega^2/\omega_{c,d}^2(\varepsilon)}
\end{align}
with a crossover frequency $\omega_{c,d}(\varepsilon)=s^2/D_d^{(\varepsilon)}$. We are interested only in the case of weak disorder $K\ll1$ and temperatures $T\gg\gamma_0$. Effective crossover frequency which enter the result for  the total attenuation, as given by Eq.(\ref{alpha_dwave_diff-e}), is
\be
	\label{dwave_crossover}
	\omega_{c,d}(T)
	=
	\frac{\pi s^2}{2v_F^2}\dfrac{ KT}{[1-K\ln(\Delta_0/T)]^3},\quad T\gg\gamma_0.
\ee
In  Eq.(\ref{dwave_crossover}) we have used  identity $K\ln(T/\gamma_0)\equiv1-K\ln(\Delta_0/T).$

At low frequencies the resulting attenuation rate has a $\propto T^2\ln T$ temperature dependence
\be
	\alpha_{d,d}(\omega)=\frac{1}{3}\frac{v_F}{v_g}\frac{T^2\omega^2}{\rho_ms^4}\ln(T/\gamma_0),\quad\omega\ll\omega_{c,d}(T),
\ee
while for higher frequencies the attenuation behaves roughly as $\propto T^4$,
\begin{align}
	\alpha_{d,d}(\omega)=\frac{7\pi^4}{60\rho_mv_F^3v_g }\frac{KT^4}{[1-K\ln(\Delta_0/T)]^5},
	\\
	\nn
	\omega\gg\omega_{c,d}(T).
\end{align}
Additional two powers of temperature, $\propto T^4$ vs. $\propto T^2$ in the normal state, are due to nearly linear DOS at high energies, $\nu_d(\varepsilon)\propto\varepsilon$ up to a logarithmic factors. Note that the energy dependence of diffusion coefficient has the same origin,  \mbox{$D_d^{(\varepsilon)}\propto\gamma(\varepsilon)\propto\nu_d^{-1}(\varepsilon)\propto\varepsilon^{-1}$}.

Contribution of energy diffusion channel is especially prominent for directions which correspond to minimum of the "local" attenuation rate (Fig.\ref{fig:dwave_angular}, Eq.\ref{alpha_dwave_local}). For such a direction at lowest frequencies, when $\omega \leq \omega_{c,d}$, the ratio of the contributions from the energy diffusion and from local channel is similar to that in the normal case,
\begin{align}
	\label{dwave_attenuation_1}
	\frac{\alpha_{d,d}}{(\alpha_{d,l})_{min}}
	=
	\frac{2\pi^2}{3}\lp\frac{v_F}{v_g}\rp^2\frac{T^2}{p_F^2s^2}.
\end{align}
For higher frequencies $\omega \gg \omega_{c,d}$ for the ratio of attenuation rates we find
\begin{align}
	\frac{\alpha_{d,d}}{(\alpha_{d,l})_{min}}
	=
	\frac{14\pi^4}{15}\lp\frac{v_F}{v_g}\rp^2\frac{T^2}{p_F^2s^2}\lp\frac{\omega_c}{\omega}\rp^2,
\end{align}
All these results are valid as long as we are in a dirty limit $ql\ll1$. This condition is equivalent to $ D_d(T)q^2\ll\gamma(T)$ that in turn gives $\omega\ll\omega_{0}^{(d)}(T)$,
\begin{align}
	\label{omega_dirty_dwave}
	\omega_{0}^{(d)}(T)\simeq\frac{s}{v_F}\frac{\gamma(T)}{\sqrt{Z(T)}}
	=
	\frac{\pi s}{2v_F}
	\frac{KT}{[1-K\ln(\Delta_0/T)]^{5/2}}.
\end{align}.

Using the parameters for  BSCCO compounds\cite{bscco1,bscco2,bscco_vratio,bscco_s}, we estimate the ratio of the (isotropic) contribution of the energy diffusion  channel (given by Eq.\ref{dwave_attenuation_1}) to the contribution of local channel at its minimum:
\begin{align}
	\frac{\alpha_{d,d}}{(\alpha_{d,l})_{min}}\approx10,\quad f=2\pi\omega\lesssim 1GHz.
\end{align}
Here we used the following values of relevant parameters: $n=5\cdot10^{21}cm^{-3}$, $v_F=2.5\cdot10^7cm/s$, $v_F/v_g=20$, $T=10K$, impurity scattering rate $\gamma_0=1K$ and sound velocity $s=4.6\cdot10^5cm/s$.

\section{Conclusions}

We have shown that low-frequency phonons in disordered conductors and superconductors experience additional damping due to coupling between lattice density  modulations and  non-equilibrium distribution of thermally excited quasiparticles. Nonequilibirum distribution of quasiparticle's energy then slowly decays leading to the phenomenon similar to the Mandelstam-Leontovich relaxation.  The effect is especially strong in doped semiconductors at moderately low $T/E_F$ and ultrasound frequencies $\omega \ll T/\hbar$.  In particular, we have estimated ultrasound attenuation in doped Si to be enhanced by factor about 100 due to this mechanism. Frequency dependence of the attenuation rate $\alpha(\omega)$ contains typical crossover frequency $\omega_c$, see Eq.(\ref{ratio1}), which depends explicitly  on electron diffusion coefficient $D$. Thus measurements of $\alpha(\omega)$ may be used for determination of $D$. Similar phenomenon exists in superconductors as well; for conventional s-wave superconductors it is weak usually, but can be important for very low-density materials close to the BCS-BEC crossover, like recently discovered YPtBi superconductor. For d-wave superconductors  we calculated both conventional (local) and diffusion-induced attenuation rate and show that the latter may dominate in strongly anisotropic BSCCO materials at moderate temperatures.

\section{Acknowledgements}

We are grateful to V.E.Kravtsov for useful discussions.
This research was supported by the RFBR grant \# 13-02-00963.

\appendix

\section{Electron-phonon vertex}

In the main text we have used the tensor-like electron-phonon vertex,
\begin{align}
	(\check{H}_{el-ph}^{CFR})_1
	=\int(d\vec r)
	\overline{\psi}
	\Big(
	p_\alpha\partial_\beta\check{H}
	- (p_Fv_F/d)\delta_{\alpha\beta}
	\Big) \partial_\beta u_\alpha
	\psi.
\end{align}
However,careful calculation shows that the actual electron-phonon interaction in a co-moving reference frame contains additional contribution\cite{us},
\begin{align}
	\check{H}_{el-ph}^{CFR}
	&=
	(\check{H}_{el-ph}^{CFR})_1+(\check{H}_{el-ph}^{CFR})_2,
	\\
	(\check{H}_{el-ph}^{CFR})_2
	&=\int(d\vec r)
	\overline{\psi}
	\Big[
	-U(\vec r)(\partial_\alpha u_\alpha)\Big]
	\psi.
\end{align}
This contribution describes changes in disorder potential \textit{strength}, for example changes in local concentration of impurities. This term is usually negligible. However, this is not the case for a d-wave state, where it plays an important role.

\subsection{Na\"{i}ve (incomplete) effective vertex}

Let us first show what would be obtained for an effective phonon-diffuson vertex ignoring the disorder-related contribution. Taking into account only $(\check{H}_{el-ph}^{CFR})_1$ we have
\be
	\Braket{\check{\Gamma}}_{d1}^{\alpha\beta}
	=
	u\sum_{nodes}
	\int(dp)\check{G}^R_+
	\lp
	p_\alpha\partial_\beta\check{H}
	- (p_Fv_F/d)\delta_{\alpha\beta}
	\rp
	\check{G}^A_-.
\ee

For longitudinal phonons, after the summation over the nodes is performed, the relevant terms arising from the vertex are
\be
	\sum_{nodes}
	\lp
	p_\alpha\partial_\beta\check{H}
	- (p_Fv_F/d)\delta_{\alpha\beta}
	\rp
	=
	2\Big[\xi\check{\tau}_0+\Delta(i\check{\tau}_2)\Big]\delta_{\alpha\beta}.
\ee
We will omit $\delta_{\alpha\beta}$ and $\alpha-,\beta-$ indices for brevity. The $\xi\check{\tau}_0$ term for example gives
\begin{align}
	\Braket{\check{\Gamma}}_{d1a}&=u\int(d\vec p)\frac{2\xi^2\tilde{\varepsilon}\check{\tau}_3}{((\varepsilon_k-\tilde{\varepsilon})^2+\gamma^2)^2}
	\\
	&=\tilde{\varepsilon}(\varepsilon)\lp1-\frac{1}{2}Z(\varepsilon)\rp\check{\tau}_3
	,
\end{align}
where $\varepsilon_k=\sqrt{\xi^2+\Delta^2}$. The same contribution arises from the $\Delta(i\check{\tau}_2)$ term thus resulting in
\be
	\label{appendix_gamma1}
	\Braket{\check{\Gamma}}_{d1}=\tilde{\varepsilon}(\varepsilon)\Big(2-Z(\varepsilon)\Big)\check{\tau_3}.
\ee
In general case the vertex Eq.$(\ref{appendix_gamma1})$ depends on electron energy in a sophisticated way. Meanwhile, in the clean limit, when the impurities are negligible and $\tilde{\varepsilon}(\varepsilon)\rightarrow\varepsilon$, $Z(\varepsilon)\rightarrow1$, we would have a very simple result $\Braket{\check{\Gamma}}_{d1}=\varepsilon\check{\tau}_3$. It turns out that this is in fact the correct answer once everything is taken into account.
\newline

\subsection{Full effective vertex}

What we have just left out is $(\check{H}_{el-ph}^{CFR})_2$, the disorder-induced contribution to the electron-phonon interaction:
\be
	\Braket{\check{\Gamma}}_{d2}^{\alpha\beta}
	=
	u\sum_{nodes}
	\int(dp)\check{G}^R_+
	\lp
	U
	\rp
	\check{G}^A_-.
\ee
After we average this vertex over the disorder we see that it gives
\begin{align}
	\Braket{\check{\Gamma}}_{d2}
	&=
	u\sum_{nodes}
	\int(dp)\check{G}^R_+
	\lp
	\check{\Sigma}^R+\check{\Sigma}^A
	\rp
	\check{G}^A_-
	\\
	\nn
	&=\dots
	\\
	\nn
	&=2\Re\check{\Sigma}^R=2\Big(\varepsilon-\tilde{\varepsilon}(\varepsilon)\Big)\check{\tau}_3.
\end{align}
In other words, this vertex is tightly connected to the renormalization of electron energy. If energy is not renormalized then this vertex is absent.

Recalling the relation $\varepsilon=Z(\varepsilon)\tilde{\varepsilon}(\varepsilon)$ we now see that the actual effective phonon-diffuson vertex is indeed connected only to the true electron energy variable $\varepsilon$:
\begin{align}
	\Braket{\check{\Gamma}}_{d1}=(2\tilde{\varepsilon}-\varepsilon)\check{\tau}_3,\quad \Braket{\check{\Gamma}}_{d2}=(2\varepsilon-2\tilde{\varepsilon})\check{\tau}_3,
\end{align}
so that
\be
	\Braket{\check{\Gamma}}_{d}
	=
	\Braket{\check{\Gamma}}_{d1}+\Braket{\check{\Gamma}}_{d2}
	=
	\varepsilon\check{\tau}_3,
\ee
exactly the expression that was used in the main text.


\begin{thebibliography}{99}
\bibitem{yptbi_s} We estimate sound velocity of YPtBi with that of a bismuth.
\end{thebibliography}
\end{document}